\documentclass{PoS}

\usepackage{bm}

\newcommand{\maybepagebreak}{\pagebreak[4]}

\PoS{PoS(LAT2005)206}

\title{Predictions from Lattice QCD \hfill
{\sf\normalsize FERMILAB-CONF-05-428-T}}

\ShortTitle{Predictions from Lattice QCD}

\author{\speaker{A.S. Kronfeld}$^{,a}$,
	I.F.~Allison$^b$,
	C.~Aubin$^{c,d}$,
	C.~Bernard$^d$,
	C.T.H.~Davies$^b$,
	C.~DeTar$^e$,
	M.~Di~Pierro$^f$,
	E.D.~Freeland$^g$,
	Steven~Gottlieb$^h$,
	A.~Gray$^i$,
	E.~Gregory$^j$,
	U.M.~Heller$^k$,
	J.E.~Hetrick$^l$,
	A.X.~El-Khadra$^m$,
	L.~Levkova$^h$,
	P.B.~Mackenzie$^a$,
	F.~Maresca$^e$,
	D.~Menscher$^m$,
	M.~Nobes$^{n,o}$,
	M.~Okamoto$^a$,
	M.B.~Oktay$^m$,
	J.~Osborn$^e$,
	D.~Renner$^j$,
	J.N.~Simone$^a$,
	R.~Sugar$^p$,
	D.~Toussaint$^j$,
	H.D. Trottier$^o$\\
        E-mail: \email{ask@fnal.gov}\\
        $^a$Fermi National Accelerator Laboratory,
	Batavia, Illinois 60510, USA\\
	$^b$Department of Physics and Astronomy, Glasgow University,  
	Glasgow, Scotland G12 8QQ, UK\\
        $^c$Physics Department, Columbia University,
	New York, New York 10027, USA\\
        $^d$Department of Physics, Washington University,
	St.~Louis, Missouri 63130, USA\\
        $^e$Physics Department, University of Utah,
	Salt Lake City, Utah 84112, USA\\
        $^f$School of Computer Science, Telecommunications and Information
	Systems, DePaul University, Chicago, Illinois 60604, USA\\
	$^g$Liberal Arts Department, School of the Art Institute,
	Chicago, Illinois 60603, USA\\
        $^h$Department of Physics, Indiana University,
	Bloomington, Indiana 47405, USA\\
        $^i$Department of Physics, The Ohio State University,
	Columbus, Ohio 43210, USA\\
        $^j$Department of Physics, University of Arizona,
	Tucson, Arizona 85721, USA\\
        $^k$American Physical Society, Ridge, New York 11961, USA\\
        $^l$Physics Department, University of the Pacific,
	Stockton, California 95211, USA\\
        $^m$Physics Department, University of Illinois,
	Urbana, Illinois 61801, USA\\
	$^n$Laboratory of Elementary-Particle Physics, Cornell University,
	Ithaca, New York 14853, USA\\
        $^o$Physics Department, Simon Fraser University,
	Burnaby, British Columbia V5A~1S6, Canada\\
        $^p$Department of Physics, University of California,
	Santa Barbara, California 93106, USA}

\author{Fermilab Lattice, MILC, and HPQCD Collaborations}

\abstract{
In the past year, we calculated with lattice QCD three quantities that
were unknown or poorly known.
They are the $q^2$ dependence of the form factor in semileptonic
$D\to Kl\nu$ decay, the decay constant of the $D$ meson, and
the mass of the $B_c$ meson.
In this talk, we summarize these calculations, with emphasis on their
(subsequent) confirmation by experiments.

}

\FullConference{XXIIIrd International Symposium on Lattice Field Theory\\
		25-30 July 2005\\
		Trinity College, Dublin, Ireland}

\begin{document}

\section{Introduction and Background}

In recent years, lattice QCD has reached the stage where many
calculations of hadron masses, mass splittings, and operator matrix
elements agree with experimental measurements.
The key has been the inclusion of sea quarks.
The progress has been especially striking~\cite{Davies:2003ik} when
the light quarks (sea and valence) are implemented as staggered quarks,
with an improved action.

Some of the ingredients of these calculations are controversial.
Staggered quarks come in four tastes, three of which must be removed
to obtain each individual flavor.
For sea quarks, this is done by taking the fourth root of the fermion
determinant;
for valence quarks, by projecting onto the desired taste sector.
Furthermore chiral perturbation theory must be modified~\cite{Aubin:2003mg}.
Although evidence for the validity of these ``tricks'' is slowly
accumulating, a proof remains at large~\cite{Durr:2005ax}.
In addition, much of the success of Ref.~\cite{Davies:2003ik} comes
from hadrons with heavy quarks.
Although debate on heavy quarks in lattice QCD seems to have subsided,
checks are still useful.

In this paper, we discuss three calculations, with emphasis on their
subsequent experimental confirmation.
They are the normalization and $q^2$-dependence of the
$D\to Kl\nu$ form factor;
the decay constants of the $D^+$ and $D_s$ mesons;
and the mass of the $B_c$ meson.
Each tests a somewhat different combination of the ingredients, and the
following table gives an informal guide:
\begin{center}
% \begin{table}[h]
% 	\centering
% 	\caption[tab:stars]{Stars.}
% 	\label{tab:stars}
	\begin{tabular}{c|ccc}
	\hline\hline
% 	calculation & $[\det_4M]^{1/4}$ & $\tr\to\frac{1}{4}\tr$ & heavy \\
	calculation & light sea & light valence & heavy \\
	\hline
	semileptonic $f_+(q^2)$ & $\star\star$ & $\star\star$ & $\star\star$ \\
	leptonic $f_D$ & $\star\star$ & $\star\star\star$ & $\star\star$ \\
	$B_c$ mass & $\star\star$ & --- & $\star\star\star$ \\
	\hline\hline
	\end{tabular}
% \end{table}
\end{center}
The chiral extrapolation, which is more sensitive to valence quarks than
sea quarks, turned out to be more important for the decay constant than
the form factor.
The $B_c$ meson has no light valence quarks at all, but one should
expect an accurate calculation only if heavy-quark discretization
effects are under control.

Successful predictions are, of course, not a substitute for a proof.
They are still useful.
Even if the experts are confident of all the elements of their numerical
calculations, non-experts are interested in an end-to-end
check~\cite{Shipsey:2004wz}.
The quantities discussed here are ideal candidates: they are
straightforward to compute; the first ``good'' experimental measurements
were not expected until this year; and new physics is unlikely to
contribute significantly.

\section{Semileptonic $\bm{D}$ Decays}

Semileptonic decays such as $D\to Kl\nu$ are mediated by electroweak
vector currents.
The matrix element $\langle K|V^\mu|D\rangle$ is parametrized by
form factors.
For a vector current there are two, but experimentally only the one
called $f_+(q^2)$ is accessible; the rate from the other one,
$f_0(q^2)$, is suppressed by~$m_l^2$.
Here $q^2$ is the momentum transferred to the lepton-neutrino system,
falling in the range $0\leq q^2\leq q^2_{\rm max}=(m_D-m_K)^2$.
In lattice QCD, discretization effects are smallest when the
momentum~$\bm{p}$ of the kaon is small,
and then $q^2$ is not too far from~$q^2_{\rm max}$.

Experiments usually measure the branching fraction and quote
the normalization $f_+(0)$, after making assumptions about the
$q^2$~dependence.
While our results were still preliminary~\cite{Okamoto:2003ur},
experimental results came out for the normalization of
$D\to Kl\nu$~\cite{Ablikim:2004ej}
and $D\to\pi l\nu$~\cite{Huang:2004fr}.
The agreement with our final results~\cite{Aubin:2004ej} is excellent.
For example, we find $f_+^{D\to K}(0)=0.73(3)(7)$~\cite{Aubin:2004ej}
while BES measures $f_+^{D\to K}(0)=0.78(5)$~\cite{Ablikim:2004ej}.
Our calculations of the normalization are also consistent with
the soft pion theorem, which states $f_0(q^2_{\rm max})=f_D/f_\pi$.

In principle, the shape of the form factors can be computed
directly in lattice QCD.
In practice, we calculated at a few values of $\bm{p}$ and used
the Be\'cirevi\'c-Kaidalov (BK) form~\cite{Becirevic:1999kt} to fix the
full $q^2$ dependence of $f_+$ and~$f_0$.
Then the normalization of $f_+$ comes mainly from $f_0$ through a
kinematic constraint $f_+(0)=f_0(0)$.
The BK Ansatz and calculations near $q^2_{\rm max}$ determine the shape.
It was important, therefore, to measure the $q^2$ dependence
experimentally.
In photoproduction of charm off fixed nuclear targets, the FOCUS
Collaboration was able to collect high enough statistics to trace out
the $q^2$ distribution of the decay~\cite{Link:2004dh}.
This setup does not yield an absolutely normalized branching ratio, so
one is left to compare $f_+(q^2)/f_+(0)$.

\begin{figure}[h]
	\centering
	\includegraphics[height=2in]{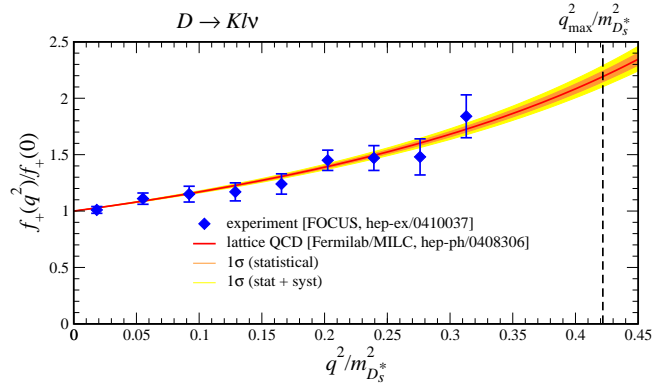}
	\caption[fig:focus]{Shape of form factor $f_+(q^2)/f_+(0)$
	vs.~$q^2/m^2_{D^*_s}$,
	compared with experiment~\cite{Link:2004dh}.}
	\label{fig:focus}
\end{figure}

In Fig.~\ref{fig:focus} we plot our result for $f_+(q^2)/f_+(0)$
vs.~$q^2/m^2_{D^*_s}$.
The errors from $f_+(0)$ must be propagated to non-zero~$q^2$,
so for $f_+(q^2)/f_+(0)$ the errors grow with~$q^2$.
Figure~\ref{fig:focus} shows 1-$\sigma$ bands of statistical (orange)
and all uncertainties (yellow) added in quadrature~\cite{Mackenzie:2005di}.
As one can see, the $q^2$ dependence of lattice QCD (curve and error
band) and experiment (points) agree excellently, although
the uncertainties are still several per cent.

\section{Leptonic $\bm{D}$ Decays}

We also computed the hadronic matrix element for the leptonic decay of
charmed mesons, $f_{D^+}$ and~$f_{D_s}$.
The first (experimental) measurements of $f_{D^+}$ appeared in~2004,
with three events from BES~\cite{Ablikim:2004ry} and
eight from CLEO~\cite{Bonvicini:2004gv}.
Neither provides a stringent test of QCD, but CLEO-$c$ was just starting
its run and promised 5--8 times higher statistics by the Summer~2005
Lepton-Photon Symposium~\cite{Shipsey:2004wz}.
At Lattice 2004 \cite{Simone:2004fr}, we presented preliminary results
for $f_{D^+}$, based on one lattice spacing, $a\approx0.125$~fm.
Our aim was to extend the running to two other lattice spacings and, of
course, to improve our understanding of other aspects of the
calculation, such as the chiral extrapolation.
Details are given in the ensuing publication~\cite{Aubin:2005ar}.
We find
\begin{equation}
	f_{D^+} = 201 \pm 3 \pm 17~\textrm{MeV},
	\label{eq:fD+:lat}
\end{equation}
where the first error is from finite Monte Carlo statistics, the second
is a sum in quadrature of several systematics.
A~conservative (but not na\"ive) estimate of heavy-quark discretizations
effects, as discussed in Ref.~\cite{Kronfeld:2003sd}, is the
second largest (largest) systematic on $f_{D^+}$ ($f_{D_s}$).
A~few days after our paper was posted on the arXiv, CLEO-$c$ announced
its new measurement~\cite{Artuso:2005lp}
\begin{equation}
	f_{D^+} = 223 \pm 17 \pm 3~\textrm{MeV},
	\label{eq:fD+:expt}
\end{equation}
based on $47\pm8$ events.
At the 1-$\sigma$ level, the agreement between Eqs.~(\ref{eq:fD+:lat})
and~(\ref{eq:fD+:expt}) is fine.
One should keep in mind that the experiment actually determines
$|V_{cd}|f_{D^+}$.
CLEO-$c$~\cite{Artuso:2005lp} assumes that $|V_{cd}|=|V_{us}|$ and
uses a recent average of~$|V_{us}|$ from semileptonic $K$ decay.

\begin{figure}[h]
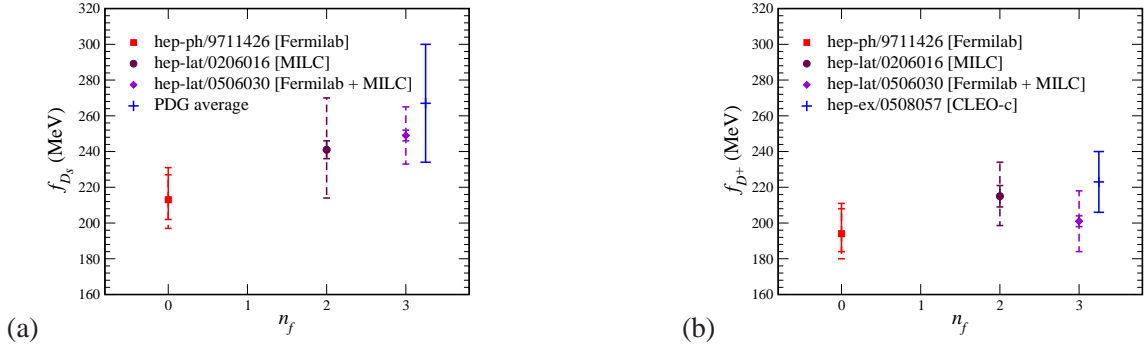

	\centering
	(a) \includegraphics[height=1.75in]{fDs-nf}\hfill
	(b) \includegraphics[height=1.75in]{fD+-nf}
	\caption[fig:mBc]{Dependence of (a) $f_{D_s}$ and (b) $f_{D^+}$
	on the number $n_f$ of sea flavors.
	Quenched ($n_f=0$)~\cite{El-Khadra:1997hq};
	$n_f=2$~\cite{Bernard:2002pc};
	$n_f=3$~\cite{Aubin:2005ar}.
	Solid (dashed) error bars are statistical (statistical+systematic).}
	\label{fig:fD}
\end{figure}

It is interesting to look at the $n_f$ dependence of $f_{D_s}$, shown in
Fig.~\ref{fig:fD}(a).
Of course, quenched results vary widely, but we show
one~\cite{El-Khadra:1997hq} carried out with similar choices for heavy
quarks, renormalization factors, etc.
One sees a trend of $f_{D_s}$ to increase with $n_f$.
A~similar comparison of $f_{D^+}$, in Fig.~\ref{fig:fD}(b), is less
instructive, because the chiral extrapolations in
Refs.~\cite{El-Khadra:1997hq,Bernard:2002pc} started at large quark
masses and are, hence, less reliable than in the present work.

\section{Mass of the $\bm{B_c}$ Meson}

The pseudoscalar $B_c^+$ meson is the lowest-lying bound state of a
charmed quark and a $b$ quark.
CDF~\cite{Abe:1998wi} first observed it during Run~I of the Tevatron
in the semileptonic decay $B_c^+\to J/\psi l^+\nu$.
During Run~II, D\O\ has confirmed the discovery in the same
mode~\cite{Cheu:2004zc}.
Because the neutrino is undetected, the mass resolution in semileptonic
modes is poor, $\pm(300$--$400)$~MeV.
Now, however, the upgraded detectors are able to reconstruct
hadronic modes, such as $B_c^+\to J/\psi\pi^+$, which give much much
better precision on~$m_{B_c}$ \cite{Anikeev:2001rk}.

At Lattice 2004 we presented results in nearly final
form~\cite{Allison:2004hy}, and posted the final results on the arXiv
in mid-November~\cite{Allison:2004be}:
\begin{equation}
	m_{B_c} = 6304 \pm 12^{+18}_{-~0}~\textrm{MeV},
	\label{eq:mBc:lat}
\end{equation}
where the last error is a rough estimate of residual heavy-quark
discretization effects.
% Soon afterwards, CDF announced a precise mass measurement (based on a
% 3-$\sigma$ evidence for observing $B_c^+\to J/\psi\pi^+$).
Soon afterwards, CDF announced a precise mass measurement.
They find~\cite{Acosta:2005us}
\begin{equation}
	m_{B_c} = 6287 \pm 5~\textrm{MeV},
	\label{eq:mBc:expt}
\end{equation}
which agrees with Eq.~(\ref{eq:mBc:lat}) at slightly more than
1-$\sigma$.

\begin{figure}[h]
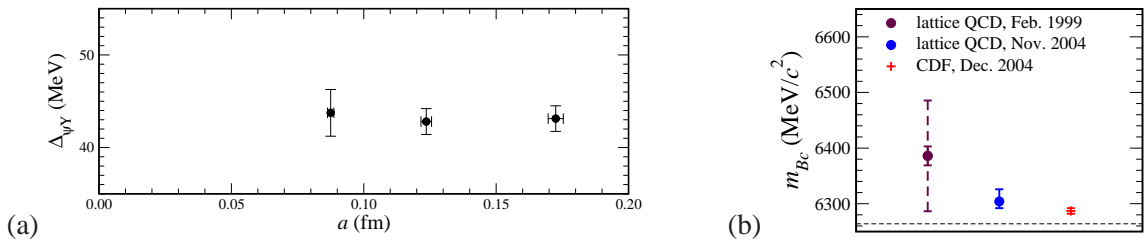

	\centering
	(a) \includegraphics[height=0.2\textwidth]{Delta-a}\hfill
	(b) \includegraphics[height=0.2\textwidth]{doe_compare}
	\caption[fig:mBc]{(a) Dependence of the
	splitting~$\Delta_{\psi\Upsilon}$ on the lattice spacing~$a$.
	(b) Comparison of the quenched~\cite{Shanahan:1999mv},
	$n_f=2+1$~\cite{Allison:2004be}, and
	experimental~\cite{Acosta:2005us} values of $m_{B_c}$;
	the dashed line denotes the baseline
	$(\bar{m}_\psi+m_\Upsilon)/2$.}
	\label{fig:mBc}
\end{figure}

Two comments are in order.
First, the agreement at the gross level of the calculation with
experiment shows that discretization effects are well under control
\maybepagebreak
with lattice NRQCD~\cite{Lepage:1987gg} and the Fermilab
method~\cite{El-Khadra:1996mp}.
Of course, this follows from the careful application of effective
field theories for heavy quarks~\cite{Lepage:1992tx,Kronfeld:2000ck}.
Indeed, as seen in Fig.~\ref{fig:mBc}(a), almost no lattice spacing
dependence is seen in the splitting
$\Delta_{\psi\Upsilon}=m_{B_c}-(\bar{m}_\psi+m_\Upsilon)/2$ that is at
the crux of the calculation~\cite{Shanahan:1999mv}.
Moreover, it is striking how much the splitting~$\Delta_{\psi\Upsilon}$
changes when sea quarks are included.
Figure~\ref{fig:mBc}(b) compares Eq.~(\ref{eq:mBc:lat}) with an old
quenched calculation~\cite{Shanahan:1999mv} (and the
measurement~\cite{Acosta:2005us}).
The solid error bar shows the non-quenching errors,
and the dashed includes the estimate of the quenching error.
The inclusion of sea quarks has reduced the splitting by a factor of
three or four, bringing an essentially discrepant result into agreement.

\section{Conclusions}

In the past year, three lattice-QCD calculations have been confirmed by
experiment.
FOCUS \cite{Link:2004dh} confirmed the $q^2$-dependence of the
$D\to Kl\nu$ form factor~\cite{Aubin:2004ej};
CLEO-$c$~\cite{Artuso:2005lp} confirmed the $D$-meson decay
constant~\cite{Aubin:2005ar};
and CDF~\cite{Acosta:2005us} confirmed the mass of the $B_c$
meson~\cite{Allison:2004be}.
To obtain these results it is essential to have heavy-quark
discretization effects under control, as one expects from theoretical
foundations~\cite{Lepage:1987gg,El-Khadra:1996mp,Lepage:1992tx,Kronfeld:2000ck}.
Furthermore, the comparison of quenched QCD, QCD with 2+1 staggered
flavors, and experiment shows that sea quarks are needed to obtain
agreement, and that staggered quarks (in these cases) capture the needed
effect.

This work has been supported in part by the U.S. National Science
Foundation, the Office of Science of the U.S. Department of Energy
(DOE), and the U.K. Particle Physics and Astronomy Research Council.
Fermilab is operated by Universities Research Association Inc., under
contract with the DOE.

\end{document}